\documentstyle[11pt]{amsart}
         
\font\twlmsbm=msbm10 scaled \magstep1
\font\egtmsbm=msbm8
\font\sixmsbm=msbm6
\newfam\msbmfam
\textfont\msbmfam=\twlmsbm
\scriptfont\msbmfam=\egtmsbm
\scriptscriptfont\msbmfam=\sixmsbm

\font\twleufm=eufm10 scaled \magstep1
\font\egteufm=eufm8
\font\sixeufm=eufm6
\newfam\eufmfam
\textfont\eufmfam=\twleufm\scriptfont\eufmfam=\egteufm
\scriptscriptfont\eufmfam=\sixeufm

\newtheorem{rem}{Remark}[section]

\newtheorem{define}{Definition}[section]

\newtheorem*{ex}{Example}
\newenvironment{example}{\begin{ex}\rm}{\end{ex}}

\def\Nabla{{\boldsymbol{\nabla}\,}}

\begin{document}
\author{N. N. Chaus}
\address{
Institute of Mathematics of Ukrainian National Academy of Sciences,	\\
vul.~ Tereshchinkivs'ka, 3,
Kiev, 252601, Ukraine,
email:chaus@@imath.kiev.ua}
\title{On the main equations of electrodynamics}

\begin{abstract}
Instead of a linear system of equations for a free electromagnetic field, we
propose a nonlinear system of equations. The classical electrodynamics is
preserved. There appeared solutions (the electromagnetic fields)
 having photon properties.The theory posits the vacuum is a physical medium. The most important problems of relativistic
interaction of interpenetrating mediums are studied.
\end{abstract}

\maketitle
\section*{The Problem}

Before a photon was experimentally discovered, the classical electrodynamics
was, at the same time, a theory of light. The locality and stability of
photons have lead specialists to a perplexity, since solutions of
electrodynamics equations do not possess these properties. For this reason,
the system of Maxwell's equations needed to be fixed by replacing it with a
nonlinear system. This problem was not solved those days. It was a new
science, the quantum electrodynamics, that should have saved the situation
and explained the phenomenon of photon. But it did not solve the problem,
and it could not solve it, for again, the same linear equations as in the
electrodynamics were used. As the result, a photon still remains nowadays an
unexplained, even mysterious object, and the theory is similar to the
astronomy of Ptolemy, accompanied by obscure philosophical doxies and
conjurations.

In this work, the author comes back to the problem of existence of such a
system of equations that would ``admit'' a photon. That is to say that the
author believes that the quantum theory is also not flawless.

\section*{Solution of the problem}

The system we are looking for comes just immediately if we will accurately
understand the main principals of electrodynamics and find correct answers
to the following questions:

\begin{itemize}
\item[a)]  what is an electric charge;

\item[b)]  what is an electric current;

\item[c)]  does an electromagnetic field act on electric charges and
currents?
\end{itemize}

\smallskip

For the reader to easier accept the author's answers to these questions,
consider a very simple example from mechanics.
\newpage

\begin{example}
Let $\phi (x)$, $-\infty <x<\infty $, be a sufficiently smooth function, and 
$\phi (x)=0$ for $|x|\geq 1$. Let us cover the graph of the function $\phi
(x)$ with a sufficiently wide plate, and cut it along the graph of the
function $\phi (x)$. Then let a stretched along the $x$-axis homogeneous
string be crimped from below and above by the parts of the cut plate, thus
making the form of the string repeat the graph of the function $\phi (x)$.
Keeping everything still let us analyze the situation. It is clear that the
deflection of the string from the initial condition $u(x)$ ($=\phi (x)$)
completely determines the state of the string. Nevertheless, let us
introduce another useful characteristic of the string. Let us call it a
string charge and set the density of the string charge to be $\sigma
(x)=\partial ^2u/\partial x^2$. It is clear that:

\begin{enumerate}
\item  The density of the force with which the lower or upper part of the
plate act on the string is proportional to the density of the string charge
(in the theory of liner string). In particular, in the places where $\sigma
(x)=0$, the string can be undercut so that it would not touch the string.
The form of the string will remain the same after such a procedure.

\item  One can consider the force with which the string acts on the plate,
and also one can consider the force with which the plate acts on the string.
One can as well consider forces acting inside the string, and forces acting
inside the plate. But saying that there are forces acting on the string
charge in the considered example is absurd. In this example the forces are
connected with the charge, these forces and the charge are interconnected,
they accompany one another, they cause one another, 
 but no forces act on the string charge. This is impossible.
\end{enumerate}
\end{example}

\smallskip

\subsection*{Answers to questions a) --- c).}

The vacuum is a physical medium. The classical electrodynamics should be
regarded first of all as a continuous theory of this medium. The known in
the electrodynamics constants $\varepsilon _0$ and $\mu _0$ are
characteristics of this medium. The fields ${\bf E}$ and ${\bf B}$ are also
main characteristics of the state of this medium.

According to the classical formula $\rho =\varepsilon _0{\rm div\,}{\bf E}$,
the author claims that the electric charge is only a special characteristic
of the state of the vacuum-medium (together with the main characteristics
--- the vectors ${\bf E}$ and ${\bf B}$).

According to the classical formula ${\bf j}=\mu _0^{-1}({\rm rot\,}%
{\bf B}-c^{-2}{\bf \dot{E}})$, the author claims that the electric current
is nothing more then another special characteristic of the vacuum-medium
(together with the main characteristics ${\bf E}$ and ${\bf B}$, and the
characteristic $\rho $).

According to the formula ${\bf f}=\rho {\bf E}+[{\bf j},{\bf B}]$ it is
accepted to think that the electromagnetic field exerts a force on currents
and charges. The author claims that there is nothing of the sort. There is
not a single experiment where one can observe an action of any force on an
electric charge. And there is not a single experiment where one could
observe an action of any force on electric current. There is not a single
physicist who could say about what exactly happens when a force acts on a
charge or current. In all experiments we only observe an action of a force
on certain physical bodies. And what acts on them is not the electromagnetic
field but the medium, the vacuum. This is similar to hydromechanics where
the action is exerted not by some characteristics of the liquid (the
distribution of speed, pressure, etc.) but acts the liquid itself, being in a
certain state.

Physical bodies on which the vacuum acts, exert a counteraction. They must.
In the continuous theory, where we are dealing not with forces but with
their densities, it is necessary and sufficient to consider an interaction
of interpenetrating mediums. The vacuum can interact simultaneously with two
distinct mediums. This is seen from the formulas $\rho =\rho ^{+}+\rho ^{-}$%
, ${\bf j}=\rho ^{+}{\bf v}^{+}+\rho ^{-}{\bf v}^{-}$. Existence of
such mediums is due to the fact that in nature there exist electrons and
protons that ``from the birth'' sit on vacuum. These particles need not be
used in the continuous theory, similarly as in the hydromechanics, the
formula $H_2O$ is not used. The mediums $M^{+}$ and $M^{-}$ do not own the
charges or currents. They only interact with the vacuum and induce a
condition where the characteristics $\rho $ and $\bf j$ appear. The
charges and currents is not a cause but a consequence. The theory about
flows of electrons, coils, magnets, capacitors, etc. is a specific part of
the theory of electrodynamics, where we are dealing with technical means to
act on vacuum, to control its state. By using recipes of this part of
electrodynamics, we prepare a special medium, or the mediums $M^{+}$ and $%
M^{-}$, as to exert a needed action on the vacuum, to get it into a needed
state. This is precisely in this part of electrodynamics where the formulas $%
{\bf j}=\rho ^{+}{\bf v}^{+}+\rho ^{-}{\bf v}^{-}$ appear that are not included
in the main equations. There are no such formulas in the theory of string;
one can touch the string with a finger, it is more difficult to reach the
vacuum.

Comparing a string and vacuum, the formulas $f=-\mu \partial _t^2u+T\partial
_x^2u$ and ${\bf f}=\rho {\bf E}+[{\bf j},{\bf B}]$ address the same
question, --- the force of an external action on the medium (the string or
vacuum). If the condition of the string, $u(x,t)$, is such that $f\equiv 0$,
then this means that the string does not interact with the ambient medium,
the string is free and self contained. The condition ${\bf f}=0$ is
necessary for the fields ${\bf E}$ and ${\bf B}$ describe the free vacuum.
But since ${\bf f}$, in the general case, is the sum of the forces ${\bf f}^{+}$
and ${\bf f}^{-}$, the condition ${\bf f}=0$ does not imply that the vacuum
is free. The mediums $M^{+}$ and $M^{-}$ could ``stretch'' the vacuum in
opposite directions and yield ${\bf f}=0$, but still there will be a
nontrivial energy exchange between the mediums. This nontriviality can be
eliminated by imposing the condition $({\bf E},{\bf j} )=0$.

On the basis of the preceding discussion the author claims that, if the
fields ${\bf E}$ and ${\bf B}$ and all other possible additional
characteristics of the vacuum condition are such that $\rho {\bf E}+[{\bf j},%
{\bf B}]=0$ and $({\bf E},{\bf j})=0$, then the vacuum is not interconnected
with anything, it is free and self contained.

\subsection*{Nonlinear system of equations}

{\em The system of equations} 
\begin{equation} \label{1}
\begin{array}{c}
\displaystyle {\bf \dot B}=-{\rm rot\,} {\bf E},\qquad {\rm div\,} {\bf B}%
=0,\qquad \rho=\varepsilon_0{\rm div\,} {\bf E}, \\[3mm] 
\displaystyle c^{-2}{\bf \dot E} + \mu_0 {\bf j}={\rm rot\,} {\bf B},\qquad
\rho{\bf E}+[{\bf j},{\bf B}]=0,\qquad ({\bf E},{\bf j})=0
\end{array}
\end{equation}
{\em is the system of equations of the state of the free vacuum\/}.

The author believes that system (\ref{1}) should replace the following
system used in physics: 
\begin{equation}
{\bf \dot{B}}=-{\rm rot\,}{\bf E},\qquad {\rm div\,}{\bf B}=0,\qquad {\rm %
div\,}{\bf E}=0,\qquad {\bf \dot{E}}=c^2{\rm rot\,}B.  \label{2}
\end{equation}

\section*{First corollaries}

If the field ${\bf E}$ (stationary or not) is sufficiently smooth and
vanishing at infinity, and the assembly $\{{\bf E};{\bf B}\equiv 0\}$ is a
solution of system (\ref{1}), then ${\bf E}\equiv 0$. Indeed, system (\ref{1}%
) yields, for ${\bf E}$, the representation ${\bf E}={\boldsymbol{\nabla}\,}
\varphi$ and $\Delta\varphi{\boldsymbol{\nabla}\,} \varphi=0$. Whence $%
\Delta\varphi=0$ and $\Delta E_i=0$ in ${\fam\msbmfam\relax R}^3$ if $%
|E_i|\to 0$ at infinity.

In the same elementary way one can deduce that there do not exist stationary
or nonstationary spherically symmetric states of the vacuum, i.e. solutions
of the form ${\bf E}={\boldsymbol{\nabla}\,} e(r,t)$, ${\bf B}={%
\boldsymbol{\nabla}\,} b(r,t)$, $r^2=x^2+y^2+z^2$.

Solutions of system (\ref{2}) satisfy system (\ref{1}). There exist
solutions of system (\ref{1}) that are not solutions of system (\ref{2}).
Let us show this.

Let $a_0(x,y,z)$ be a sufficiently smooth function on ${\fam\msbmfam\relax R}%
^3$ with compact support. Denote by $a=a_0(x-ct,y,z)$ and let ${\bf E}%
=(0,c\partial _ya,c\partial _za)$, ${\bf B}=(0,-\partial _za,\partial _ya)$.
Such ${\bf E}$ and ${\bf B}$ satisfy system (\ref{1}) but not (\ref{2}).

Let us stress on the following properties of the solution ${\bf E}$ and $%
{\bf B}$:

\begin{enumerate}
\item  The vectors ${\bf E}$ and ${\bf B}$ and the support of the function $%
a_0$ travel with the velocity of light along the $x$-axis without change.

\item  The vectors ${\bf E}$ and ${\bf B}$ are orthogonal to the direction
of the travel.

\item  The characteristic $\rho $ for this solution equals $\varepsilon
_0c(\partial _y^2a+\partial _z^2a)$, and, consequently, the full charge
transported by the wave is zero.

\item  The total energy ${\cal E}$ and the total momentum ${\bf P}$ of the
wave satisfy ${\cal E}=cP$.

\item  It is easy to see that, if the wave meets a similar wave with another direction
of the travel, one   
observes an interaction, since there is no superposition for system (\ref{1}%
).
\end{enumerate}

\subsection*{Special example}

Take $a_0=(Ay\sin \omega x+Bz\cos \omega x)\chi (x,y,z)$, where $\chi$
 is a sufficiently smooth function equal to zero outside of a
compact set $G_0$, and $\chi \equiv 1$ in a smaller domain $G_1$, $G_1\subset
G_0$. For such $a_0$, the solution ${\bf E}$, ${\bf B}$ in the domain $G_1$
gives a classical ellipse polarized field.

\subsection*{Hypothesis}

The photon, considered as a real object, is characterized in the classical
chlectrodynamics by the fields ${\bf E}$ and ${\bf B}$ from the preceding
construction. It is a special state of the free vacuum. The diversity of
photons is limited by a special class of the function $a_0$.

\section*{Some problems}

\subsection*{1.}

To formulate conditions on the functions $a_0$ such that the functions would
correspond to real photons. It can happen that the special example of the
considered $a_0$ serves as a useful remark on this question. Or this is not
true. The author does not understand very well how an elliptically polarized
wave could manage to go through a stationary plate of a polaroid. At this
time the author does not understand this phenomenon at all.

\subsection*{2.}

The class of functions $a_0$ corresponding to real photons will be fairly
small. It is not clear in principle whether or not it is technologically possible to
produce artificial photons.

\subsection*{3.}

The majority of problems in the electrical and radio technology can be
solved by using classical theory of electromagnetic waves that dissipate at
infinity. But as far as the real radiation is concerned, a dominating
opinion is that everything consists of photons. System (\ref{1}) contains
both types of solutions, and, hence, there is a suspicion that this is what
indeed happens in reality.

\subsection*{4.}

Let us look for solutions of system (\ref{1}) in the case where there is a
symmetry axis.

\subsubsection*{a)}

Let us consider the fields as follows: 
\begin{equation*}
{\bf E}={\boldsymbol{\nabla}\,} f(u),\qquad {\bf B}=\left (\frac xs\partial
_zg(u) + \frac ysh(u),\frac ys\partial _z g(u) -\frac xsh(u),-2\partial
_sg(u)\right ),
\end{equation*}
where $f(\lambda)$, $g(\lambda)$, $h(\lambda)$ are certain functions of one
variable, $u=u(s,z)$, $s=x^2+y^2$. For such kind of  field, we
immediately have that ${\rm rot\,} {\bf E}=0$, ${\rm div\,} {\bf B}=0$, $(%
{\bf E},{\bf j})=0$, and 
\begin{equation*}
[{\bf j},{\bf B}] =-\frac 1{\mu_0} \left (4g^{\prime}(u) \partial _s^2 g(u)
+ \frac 1s g^{\prime}(u) \partial _z^2 g(u) + \frac 1s h(u)
h^{\prime}(u)\right ){\boldsymbol{\nabla}\,} u.
\end{equation*}
To obtain this formula, we use that $\partial _x=2x\partial _s$, $\partial
_y=2y\partial _s$, $\partial _x^2+\partial _y^2=4\partial _s(s\partial _s)$.
As we see, the equation $\rho{\bf E} + [{\bf j},{\bf B}]=0$ in system (\ref
{1}) means that there is the following relation between the functions $f$, $%
g $, $h$, $u$: 
\begin{equation}  \label{*}
\frac s{c^2}\Delta f(u)\cdot f^{\prime}(u)=4sg^{\prime}(u)\partial _s^2 g(u)
+ g^{\prime}(u) \partial _z^2 g(u) + h(u) h^{\prime}(u).
\end{equation}
Formally, each collection of functions $f$, $g$, $h$, $u$ satisfying
equation (\ref{*}) generates fields ${\bf E}$ and ${\bf B}$ that are
solutions of system (\ref{1}).

Let us set in (\ref{*}) $f(\lambda)=a_0\lambda$, $g(\lambda) =a\lambda$,
$h(\lambda)=b\lambda$, where $%
a_0,a,b$ are constants. Equation (\ref{*}) becomes the linear equation for $u$%
: 
\begin{equation}  \label{**}
\frac {a_0^2}{c^2} s\Delta u = 4a^2 s \partial _s^2 u+ a^2 \partial _z^2 u +
b^2 u.
\end{equation}
Formally, each solution $u$ of equation (\ref{**}) generates a solution of
system (\ref{1}), $\displaystyle {\bf E}=a_0{\boldsymbol{\nabla}\,} u$, $%
\displaystyle {\bf B}=\left (\frac {ax}s \partial _z u + \frac {by}s u, \frac {ay}s
\partial _z u - \frac {bx}s u, -2a\partial _s u\right )$. In particular, the
function $u_k=s^k(s+z^2)^{-2k-1/2}$ satisfies the equation $s\Delta u_k=4k^2
u_k$, and, thus, the corresponding fields ${\bf E}^{(k)}={\boldsymbol{\nabla}%
\,} u_k$, $\displaystyle {\bf B}^{(k)} =\frac {2k}{cs}(yu_k,-xu_k,0)$ is a
solution of system (\ref{1}), however, not in the whole space. All these
solutions $\{{\bf E}^{(k)}, {\bf B}^{(k)}\}$, $k\in
{\fam\msbmfam\relax R}^1$,
have a singularity at the origin, and hence will not serve as states of the
free vacuum. But these solutions could be of interest if considered as
supplementing the well known electromagnetic field of a stationary point
charge ($k=0$).

It can happen that there are no solutions of equation (\ref{**}) that are
regular and vanishing at infinity for $s+z^2\to \infty $. Let us explain
this. First of all if such solutions exist, then there would be sufficiently
many of them, since, together with a solution $u(s,z)$, the functions $%
u(s,z+\lambda )$, $\int_a^bu(s,z+\lambda )\varphi (\lambda )\,d\sigma
(\lambda )$, and $L(\partial _z)u(s,z)$ would also be solutions, where $%
L(\partial _z)$ is a linear polynomial in $\partial _z$ with constant
coefficients. Let now $u_1(s,z)$ and $u_2(s,z)$ be two such solutions, and $%
\{{\bf E}^1,{\bf B}^1\}$ and $\{{\bf E}^2,{\bf B}^2\}$ --- the corresponding
electromagnetic fields. Then the superposition of these fields, $\{{\bf E},%
{\bf B}\}=\{{\bf E}^1+{\bf E}^2,{\bf B}^1+{\bf B}^2\}$ also satisfies system
(\ref{1}), because $\{{\bf E},{\bf B}\}$ is generated by the solution $%
(u_1+u_2)$ of equation (\ref{**}). As far as particles are concerned, it is
apparent that particles 1 and 2 are noninteracting. After this, let us
consider the electromagnetic field $\{{\bf \tilde{E}}^2,{\bf \tilde{B}}^2\}$
which is obtained by a translation of the field $\{{\bf E}^2,{\bf B}^2\}$ in 
${\fam\msbmfam\relax R}^3$ but not along the $z$-axis. This field will also
be a solution of system (\ref{1}), but it will be generated by a solution $%
\tilde{u}_2(x,y,z)$ of a linear equation distinct from equation (\ref{**}).
As a result, the superposition $\{{\bf E}^1+{\bf \tilde{E}}^2,{\bf B}^1+{\bf 
\tilde{B}}^2\}$, in general, will not satisfy system (\ref{1}), and we get
that ``threaded on an axis'' particles do not interact, whereas they become
interacting in another position. It seems to the author that such physics is
too exotic, and thus we can leave the problem of finding solutions of
equation (\ref{**}) and again look for the functions $f$, $g$, $h$, $u$
which would satisfy the nonlinear equation (\ref{*}) and generate reasonable
electromagnetic fields.

\subsubsection*{b)}

Let a solution of system (\ref{1}) have the form: ${\bf E}=(\partial
_x\Psi,\partial _y\Psi,\partial _z\Psi+\dot\Phi) $, ${\bf B}=(-\partial _y\Phi, \partial
_x\Phi,0)$, $\Psi=\Psi(s,z,t)$, $\Phi=\Phi(s,z,t)$, $s=x^2+y^2$.

One can check that such ${\bf E}$ and ${\bf B}$ are solutions of system (\ref
{1}) if $\Psi$ and $\Phi$ satisfy the following system: 
\begin{equation*}
\begin{array}{c}
\partial_s \Phi[4c^2\partial _s(s\partial _s\Phi)-\ddot\Phi-\partial _z\dot
\Psi]=\partial _s\Psi[\Delta\Psi+\partial _z \dot \Phi], \\[3mm] 
\partial _s\Phi[4c^2\partial _z(s\partial _s\Phi)+4s\partial _s\dot \Psi%
]=(\dot \Phi+\partial _z\Psi) [\Delta\Psi+\partial _z\dot \Phi].
\end{array}
\end{equation*}
Assuming that $\Phi$ and $\Psi$ are independent of $t$ and denoting $%
s\partial _s\Phi=g$, we get that the pair $\Psi$, $g$ satisfy the system 
\begin{equation*}
4c^2 g\partial _s g=s\Delta\Psi\partial _s\Psi,\qquad 4c^2g\partial
_zg=s\Delta\Psi\partial _z\Psi.
\end{equation*}
Formally, each solution $g$, $\Psi$ of this system generates a solution of
system (\ref{1}), ${\bf E}=\Nabla\Psi$, ${\bf B}=(-2ys^{-1}g, 2xs^{-1}g,0)$%
. For example, such is the pair $\Psi$, $g=c^{-1}s\partial _s\Psi$ for an
arbitrary but independent of $z$ function $\Psi$.

The author does not have a more interesting example of a pair $g$, $\Psi$ as
well as a regular solution of equation (\ref{*}).

\subsection*{5.}

Talking about the interaction of photon solutions as they meet, we should
add that there are no known formulas describing interactions between photon
solutions and a stationary electromagnetic field. Only superposition of the
solution ${\bf E}=(0,c\partial _y a,c\partial _z a)$, ${\bf B}=(0,-\partial
_z a,\partial _y a)$ and a stationary field of the form ${\bf \tilde E}%
=(0,ch_2,ch_3)$, ${\bf \tilde B}=(h_1,-h_3,h_2)$ yields again a solution of
system (\ref{1}).

\section*{General theory. Necessity.}

Interaction between the vacuum and physical bodies in the continuous theory,
when the force density is used, could only be interactions between mediums
that interpenetrate each other. In the general case, we will be considering
a medium $M$ and a medium $\Phi $ that simultaneously fill a certain part of space
and have there, and in other part of the space, the velocities $v_\alpha
^M(x,y,z,t)$ and $v_\alpha ^\Phi (x,y,z,t)$, respectively. Let there be a
force interaction between the mediums. Let $f_\alpha ^M$ be the force
density with which the medium $M $ acts on the medium $\Phi$, and
let $f_\alpha^\Phi$ be the force density with which the medium $\Phi$ acts on $M$.
There is no
relation $f_\alpha ^M=-f_\alpha ^\Phi $ in the relativistic theory. It is
replaced by a more complicated formula obtained by switching from the
densities $f_\alpha ^M$, $f_\alpha ^\Phi $ to the corresponding four
dimensional force densities. The author did not succeed in obtaining a
unique transition, hence in the sequel we give two versions of all main
formulas. The reason for this is that, in every point where $f_\alpha ^M$
and $f_\alpha ^\Phi $ are not equal to zero, there are two velocities $%
v_\alpha ^M$ and $v_\alpha ^\Phi $, instead of one, that makes the forth
component of the four dimensional force density. It turns out that each one
of these velocities is  capable to control the forth component of the
4-density of either force.

To make it less confusing, we preserve the notations $f_\alpha^M$ and $%
f_\alpha^\Phi$ for the first version, and assume that the corresponding
4-densities $f_k^M$, $f_k^\Phi$ are of the form $\{f_\alpha^M,\frac ic
f_\beta^M v_\beta^M\}$ and $\{f_\alpha^\Phi, \frac ic f_\beta^\Phi
v_\beta^\Phi\}$.\footnote{%
Here and in the sequel, the main $4$-vector is $(x_1,x_2,x_3,x_4)=(x,y,z,ict)$ and
the $4$-velocity is $V_k=(\gamma v_\alpha,ic\gamma)$, $\partial _k=\partial
/\partial x_k$.}
For the second version, $g_\alpha^M$ and $g_\alpha^\Phi$
will denote the force densities exerting by $M$ onto $\Phi$ and $\Phi$ onto $%
M$, and the corresponding densities $g_k^M$ and $g_k^\Phi$ are of the form $%
\{g_\alpha^M, \frac ic g_\beta^M v_\beta^\Phi\}$ and $\{g_\alpha^\Phi, \frac
ic g_\beta^\Phi v_\beta^M\}$.

\subsection*{The formulas connecting $f_k^M$ and $f_k^\Phi$ ($g_k^M$ and $%
g_k^\Phi$)}

Denote by $V_k^M$ and $V_k^\Phi$ the $4$-velocities (fields) of the
mediums $M$ and $\Phi$, i.e. $V_\alpha^M=\gamma^M v_\alpha^M$, $%
V_\alpha^\Phi=\gamma^\Phi v_\alpha^\Phi$, $\alpha=1,2,3$, $V_4^M=ic\gamma^M$%
, $V_4^\Phi=ic\gamma^\Phi$. Denote by ${\overset{\circ }{v}}_\alpha(x,y,z,t)$
such a velocity in the initial inertial reference frame (IRF) for the new IRF' such that
the mediums $M$ and $\Phi$ would have the velocities $(v_\alpha^M)^{%
\prime}=-(v_\alpha^\Phi)^{\prime}$ at a point $(x,y,z,t)^{\prime}$ in this
new IRF'. Such ${\overset{\circ }{v}}_\alpha$ is defined by the
formula ${\overset{\circ }{v}}_\alpha=(\gamma^M+\gamma^\Phi)^{-1}(V_%
\alpha^M+V_\alpha^\Phi)$, the corresponding $4$-velocity ${\overset{\circ }{V%
}}_k$ will be $({\overset{\circ }{\gamma }} {\overset{\circ }{v}}_\alpha,ic{%
\overset{\circ }{\gamma}})$ with ${\overset{\circ }{\gamma}}%
=(\gamma^M+\gamma^\Phi)(2+ 2 \gamma^M \gamma^\Phi - 2 \gamma^M \gamma^\Phi
c^{-2} v_\alpha^M v_\alpha^\Phi)^{-1/2}$. It is natural to think that
IRF' have the property that observed forces that act and counteract
between the mediums at this very point $(x,y,z,t)^{\prime}$ differ by the
sign, i.e. 
\begin{equation*}
\begin{array}{c}
(f_\alpha^M)^{\prime}=-(f_\alpha^\Phi)^{\prime},\quad
(g_\alpha^M)^{\prime}=-(g_\alpha^\Phi)^{\prime},\quad \alpha=1,2,3, \\[3mm] 
(f_4^M)^{\prime}=(f_4^\Phi)^{\prime},\quad
(g_4^M)^{\prime}=(g_4^\Phi)^{\prime}.
\end{array}
\end{equation*}
We should also add that all the densities with and without prime are
connected by the Lorenz transformation defined by the velocity ${\overset{%
\circ }{v}}_\alpha$. All this leads to the following formulas for the
initial IRF ($k,m=1,\dots ,4$): 
\begin{eqnarray}
f_k^\Phi= -f_k^M-\frac 2{c^2} f_m^M {\overset{\circ }{V}}_m {\overset{\circ 
}{V}}_k, \qquad f_k^M {\overset{\circ }{V}}_k = f_k^\Phi {\overset{\circ }{V}%
}_k,  \label{3} \\
g_k^\Phi=-g_k^M-\frac 2{c^2} g_m^M {\overset{\circ }{V}}_m {\overset{\circ }{%
V}}_k,\qquad g_k^M {\overset{\circ }{V}}_k=g_k^\Phi {\overset{\circ }{V}}_k.
\label{4}
\end{eqnarray}
A boring derivation of these formulas are left to the reader.

\subsection*{The medium energy-momentum tensors}

The interacting mediums $M$ and $\Phi $ have some domain $G=G_M\cap G_\Phi
\subset {\fam\msbmfam\relax R}^3$ of their mutual existence as well as
regions in ${\fam\msbmfam\relax R}^3$ where they exist by themselves. This
fact suggests that one must define the energy and write its conservation law
separately for each medium counting the energy exchange and the energy
transformation from one form into another. We thus assume that the variables 
$f_4^M$ and $f_4^\Phi $ (or $g_4^M$ and $g_4^\Phi $) together determine the
energy of both the energy of the medium $M$ and the energy of the medium $%
\Phi $, and the conservation laws have the form: 
\begin{eqnarray}
\dot{W}^M+{\rm div\,}{\bf S}^M &=&-k_Mf_\alpha ^Mv_\alpha ^M+k_\Phi f_\alpha
^\Phi v_\alpha ^\Phi ,  \label{5} \\
\dot{W}^\Phi +{\rm div\,}{\bf S}^\Phi &=&-k_\Phi f_\alpha ^\Phi v_\alpha
^\Phi +k_Mf_\alpha ^Mv_\alpha ^M,  \label{6}
\end{eqnarray}
or 
\begin{eqnarray}
\dot{\widetilde{W}}^M+{\rm div\,}{\bf \tilde{S}}^M &=&-\kappa _\Phi g_\alpha
^Mv_\alpha ^\Phi +\kappa _\Phi g_\alpha ^\Phi v_\alpha ^M,  \label{7} \\
\dot{\widetilde{W}}^\Phi +{\rm div\,}{\bf \tilde{S}}^\Phi &=&-\kappa _\Phi
g_\alpha ^\Phi v_\alpha ^M+\kappa _Mg_\alpha ^Mv_\alpha ^\Phi .  \label{8}
\end{eqnarray}
Here $W^M$, $W^\Phi $ (or $\widetilde{W}^M$, $\widetilde{W}^\Phi $) are energy
densities in $M$ and $\Phi $, ${\bf S}^M$, ${\bf S}^\Phi $ (or ${\bf \tilde{S%
}}^M$, ${\bf\tilde{S}}^\Phi $) are energy fluxes, if the forces $f_\alpha ^M$, $%
f_\alpha ^\Phi $ (or $g_\alpha ^M$, $g_\alpha ^\Phi $) operate. The
parameters $k_M$, $k_\Phi $, $\kappa _M$, $\kappa _\Phi $ control the energy
exchange between the mediums, i.e. these parameters make a quantitative
characteristic of the pair of mediums. We assume that they are all positive
and $k_M+k_\Phi =\kappa _M+\kappa _\Phi =1$. The case where $k_M=k_\Phi $ ($%
\kappa _M=\kappa _\Phi $) correspond to a symmetric interaction between the
mediums $M$ and $\Phi $. Note that each of the equations (\ref{5}) --- (\ref
{8}) is considered in its own domain ($G_M$ or $G_\Phi $).

Starting with formulas (\ref{5}) --- (\ref{8}) and using the $4$-vectors $%
f_k $ and $g_k$ we introduce the tensors $\tau_{ik}^M$, $\tau_{ik}^\Phi$, $%
\tilde \tau_{ik}^M$, $\tilde \tau_{ik}^\Phi$ such that the following
equations hold (in the corresponding domains):
$$
\partial _k \tau_{ik}^M = k_M f_i^M - k_\Phi f_i^\Phi,\qquad \partial_k
\tau_{ik}^\Phi = k_\Phi f_i^\Phi - k_M f_i^M, \eqno {(11),\ (12)}
$$
$$
\partial _k \tilde \tau_{ik}^M = \kappa_M g_i^M - \kappa_\Phi
g_i^\Phi,\qquad \partial _k \tilde \tau_{ik}^\Phi = \kappa_\Phi g_i^\Phi -
\kappa_M g_i^M. \eqno {(13),\ (14)}
$$
\setcounter{equation}{14} The signs in these equations are put in such a way
that $\tau_{44}^M$, $\tau_{44}^\Phi$, $\tilde \tau_{44}^M$, $\tilde
\tau_{44}^\Phi$ could serve in equations (\ref{5}) --- (\ref{8}) as energy
densities, and the vectors $ic\tau_{4\alpha}^M$, $ic\tau_{4\alpha}^\Phi$, $%
ic\tilde \tau_{4\alpha}^M$, $ic\tilde \tau_{4\alpha}^\Phi$ --- the energy
fluxes. The tensors defined by equations (\ref{5}) --- (14) could be called
energy-momentum tensors for the mediums $M$ and $\Phi$. Also the vectors $%
ic^{-1} \tau_{\alpha4}^M$, $ic^{-1} \tau_{\alpha4}^\Phi$, $ic^{-1} \tilde
\tau_{\alpha4}^M$, $ic^{-1} \tilde \tau_{\alpha4}^\Phi$ will be impulse
densities for the mediums $M$ and $\Phi$.

If one of the parameters $k_M$, $k_\Phi $, $\kappa _M$, $\kappa _\Phi $
equals $0$ or $1$, then this is a case of a limiting nonsymmetrical
interaction between the mediums. In particular, if $k_M=1$ and $k_\Phi =0$,
we get from (11), (12) that 
\begin{equation*}
\partial _k\tau _{ik}^MV_i^M=0,\qquad \partial _k\tau _{ik}^\Phi V_i^M=0.
\end{equation*}
The first of these conditions appears, for example, in relativistic
hydrodynamics. There $V_i^M$ is a field of $4$-velocities of the liquid, and
the tensor $\tau _{ik}^M$ is the energy-momentum tensor of the liquid itself
that interacts with the so-called mass-forces (a one more phantom). The
second condition is fundamental in electrodynamics. There $\tau _{ik}^\Phi $
is the Poynting's energy-impulse tensor which characterizes the state
of the vacuum, $V_i^M$ is not a $4$-velocity of the vacuum but that of the
medium $M$ interacting with the vacuum $\Phi $. Existence in physics of
these two very different formulas on a similar subject lead the author to an
understanding that, in the general theory of interacting mediums, there must
appear special control parameters $k_M$, $k_\Phi $, $\kappa _M$, $\kappa
_\Phi $. It is clear that the energy states of both mediums could influence
to a great extend the process of energy exchange between the mediums, and,
hence, the scalars $k_M$, $k_\Phi $, $\kappa _M$, $\kappa _\Phi $, in
general, are not constants but, for example, if a process is considered in a
small volume and for a short period of time, could be regarded as such.

By using equations (\ref{5}) --- (14) one can obtain a series of other
equations eliminating some density from (\ref{5}) --- (14) by using formulas
(\ref{3}), (\ref{4}). In particular, we have 
\begin{equation}  \label{13}
\partial _k \tau_{ik}^\Phi=-f_i^M -k_\Phi \frac 2{c^2} f_m^M {\overset{\circ 
}{V}}_m {\overset{\circ }{V}}_i,\qquad \partial \tilde \tau_{ik}^\Phi =
g_i^\Phi + \kappa_M \frac 2{c^2} g_m^\Phi {\overset{\circ }{V}}_m {\overset{%
\circ }{V}}_i.
\end{equation}

\subsection*{A more complicated interaction of mediums}

A more complicated scheme of interactions will be used to consider
electromagnetic phenomena. At this point we leave out the question on
whether an electron is a free state of the vacuum. Our goal is to obtain
analogues of formulas (\ref{13}) for three mediums $M1$, $M2$ and $\Phi $
which simultaneously occupy the same region in space $G=G_1\cap G_2\cap
G_\Phi $. This means that we have at our disposal the velocities ${\bf v}^1$%
, ${\bf v}^2$, ${\bf v}^\Phi $ and force densities ${\bf f}^{12}$, ${\bf f}%
^{21}$, ${\bf f}^{1\Phi }$, ${\bf f}^{\Phi 1}$, ${\bf f}^{2\Phi }$, ${\bf f}^{\Phi
2}$, where ${\bf f}^{12}$ is the force density with which the medium $M1$
acts on the medium $M2$, etc. Passing to $4$-densities we again obtain 2
versions of them: $f_k^{12}$, $f_k^{21}$, $f_k^{\Phi 1}$, $f_k^{\Phi 2}$, $%
f_k^{2\Phi }$, and $g_k^{12}$, $g_k^{12}$, $g_k^{21}$, $g_k^{\Phi 1}$, $%
g_k^{\Phi 2}$, $g_k^{2\Phi }$. Also the forces of action and counteraction
are connected by formulas similar to (\ref{3}), (\ref{4}). For example, 
\begin{eqnarray}
f_k^{\Phi 1} &=&-f_k^{1\Phi }-\frac 2{c^2}f_m^{1\Phi }{\overset{\circ }{V}}%
_m^{1\Phi }{\overset{\circ }{V}}_k^{1\Phi },\qquad f_k^{1\Phi }{\overset{%
\circ }{V}}_k^{1\Phi }=f_k^{\Phi 1}{\overset{\circ }{V}}_k^{1\Phi },
\label{14} \\
g_k^{\Phi 1} &=&-g_k^{1\Phi }-\frac 2{c^2}g_m^{1\Phi }{\overset{\circ }{V}}%
_m^{1\Phi }{\overset{\circ }{V}}_k^{1\Phi },\qquad g_k^{1\Phi }{\overset{%
\circ }{V}}_k^{1\Phi }=g_k^{\Phi 1}{\overset{\circ }{V}}_k^{1\Phi },
\label{15}
\end{eqnarray}
where the $4$-velocity ${\overset{\circ }{V}}_k^{1\Phi }={\overset{\circ }{V}%
}_k^{\Phi 1}$ is determined by using the well known procedure applied to the
pair of velocities ${\bf v}^1$, ${\bf v}^\Phi $.

There is an exchange of energies in the mediums $M1$, $M2$, $\Phi $. Let $%
W^1 $, $W^2$, $W^\Phi $ be energy densities of the mediums $M1$, $M2$, $\Phi 
$, and ${\bf S}^1$, ${\bf S}^2$, ${\bf S}^\Phi $ be flux of these energies.
The the most simple equations that could control the energy exchange between
the mediums are the following natural generalizations of (\ref{5}), (\ref{6}%
): 
\begin{eqnarray}
\dot{W}^1+{\rm div\,}{\bf S}^1 &=&-k_{1\Phi }{\bf f}^{1\Phi }{\bf v}%
^1+k_{\Phi 1}{\bf f}^{\Phi 1}{\bf v}^\Phi -k_{12}{\bf f}^{12}{\bf v}^1+k_{21}%
{\bf f}^{21}{\bf v}^2,  \label{16} \\
\dot{W}^2+{\rm div\,}{\bf S}^2 &=&-k_{2\Phi }{\bf f}^{2\Phi }{\bf v}%
^2+k_{\Phi 2}{\bf f}^{\Phi 2}{\bf v}^\Phi -k_{21}{\bf f}^{21}{\bf v}^2+k_{12}%
{\bf f}^{12}{\bf v}^1,  \label{17} \\
\dot{W}^\Phi +{\rm div\,}{\bf S}^\Phi &=&-k_{\Phi 1}{\bf f}^{\Phi 1}{\bf v}%
^\Phi +k_{1\Phi }{\bf f}^{1\Phi }{\bf v}^1-k_{\Phi 2}{\bf f}^{\Phi 2}{\bf v}%
^\Phi +k_{2\Phi }{\bf f}^{2\Phi }{\bf v}^2,  \label{18}
\end{eqnarray}
and analogous three equations, corresponding to (\ref{7}), (\ref{8}) with
the forces $g$ and constants $\kappa _{\alpha \beta }$. Similarly to the
case of two mediums, all the coefficients now are in pairs in the sense that 
$k_{1\Phi }+k_{\Phi 1}=1$, $k_{\Phi 2}+k_{2\Phi }=1$, $\kappa _{1\Phi
}+\kappa _{\Phi 1}=1$, $\kappa _{2\Phi }+\kappa _{\Phi 2}=1$, $%
k_{12}+k_{21}=1$, $\kappa _{12}+\kappa _{21}=1$.

The three formulas we gave, (\ref{16}), (\ref{17}), (\ref{18}), and the
other three formulas can be used to introduce the tensors $\tau^1_{ik}$, $%
\tau^2_{ik}$, $\tau^\Phi_{ik}$, $\tilde \tau^1_{ik}$, $\tilde \tau^2_{ik}$, $%
\tilde \tau^\Phi_{ik}$ relating them to the $4$-densities $f_k^{\alpha\beta}$
and $g_k^{\alpha\beta}$. As an example we give two such equations that
correspond to equations (11) and (14): 
\begin{eqnarray}
\partial _k \tau_{ik}^\Phi=k_{\Phi1} f_k^{\Phi1} - k_{1\Phi} f_k^{1\Phi} +
k_{\Phi2} f^{\Phi2}_k -k_{2\Phi} f_k^{2\Phi},  \label{19} \\
\partial _k \tilde \tau_{ik}^\Phi=\kappa_{\Phi1} g_k^{\Phi1} -
\kappa_{1\Phi} g_k^{1\Phi} + \kappa_{\Phi2} g^{\Phi2}_k -\kappa_{2\Phi}
g_k^{2\Phi}.  \label{20}
\end{eqnarray}
Each of the tensors $\tau_{ik}^\Phi$, $\tilde \tau_{ik}^\Phi$ could be
called an energy-momentum tensor of the medium $\Phi$. Let us replace in (21), (22) $%
f_k^{\Phi1}$ and $f_k^{\Phi2}$ by using formulas (\ref{14}), and $%
g_k^{1\Phi} $ and $g_k^{2\Phi}$ --- according to (\ref{15}). In addition
also assume that the interactions between the medium $\Phi$ and the mediums $%
M1$ and $M2$ are the same: $k_{\Phi1} = k_{\Phi2} = k_\Phi$, $%
k_{1\Phi}=k_{2\Phi}=k_M$, $\kappa_{\Phi1} = \kappa_{\Phi2} = \kappa_\Phi$, $%
\kappa_{1\Phi} = \kappa_{2\Phi} = \kappa_M$. All this leads to the following
generalization of formulas (\ref{13}): 
\begin{eqnarray}
\partial _k \tau_{ik}^\Phi = -(f_i^{1\Phi} + f_i^{2\Phi}) - k_\Phi \frac
2{c^2} ( f_m^{1\Phi} {\overset{\circ }{V}}_m^{1\Phi} {\overset{\circ }{V}}%
_i^{1\Phi} + f_m^{2\Phi} {\overset{\circ }{V}}_m^{2\Phi} {\overset{\circ }{V}%
}_i^{2\Phi}),  \label{21} \\
\partial _k \tilde \tau_{ik}^\Phi = (g_i^{\Phi1} + g_i^{\Phi2}) + \kappa_M
\frac 2{c^2} ( g_m^{\Phi1} {\overset{\circ }{V}}_m^{\Phi1} {\overset{\circ }{%
V}}_i^{\Phi1} + g_m^{\Phi2} {\overset{\circ }{V}}_m^{\Phi2} {\overset{\circ 
}{V}}_i^{\Phi2}).  \label{22}
\end{eqnarray}

\subsection*{A look at electrodynamics}

We start with the formulas $\rho =\rho ^{+}+\rho ^{-}$, $\rho ^{+}\geq 0$, $%
\rho ^{-}\leq 0$, ${\bf j}=\rho ^{+}{\bf v}^{+}+\rho ^{-}{\bf v}^{-}$ which
allow to state that the vacuum $\Phi $ interacts with a medium $M1$ that has
velocity ${\bf v}={\bf v}^{+}$, and with a medium $M2$ that has velocity $%
{\bf v}^2={\bf v}^{-}$. We now take the representation of the Lorentz $4$%
-force $f_k=F_k^{+}+F_k^{-}$, $F_k^{\pm }=\{\rho ^{\pm }{\bf E}+\rho ^{\pm }[{\bf v%
}^{\pm },{\bf B}],\frac ic\rho ^{\pm }({\bf E},{\bf v}^{\pm })\}$, and
address the question on what the $4$-forces $F_k^{+}$ and $F_k^{-}$ are and what
is their place in the theory of interaction of mediums? Comparing $F_k^{+}$
and $F_k^{-}$ with the $4$-vectors $f_k^{1\Phi }$, $f_k^{2\Phi }$, $%
g_k^{1\Phi }$, etc. we come to the conclusion that there are two answers to the
posed question: a) $F_k^{+}=g_k^{\Phi 1}$, $F_k^{-}=g_k^{\Phi 2}$, b) $%
-F_k^{+}=f_k^{1\Phi }$, $-F_k^{-}=f_k^{2\Phi }$. The answer a) reiterates
the well established notion of ${\bf f}$, the answer b) is new, and the
author sees no reasons why the answer a) is more preferable. One also must
retain both formulas (\ref{21}) and (\ref{22}) which now become: 
\begin{equation}
\begin{array}{c}
\partial _k\tau _{ik}^\Phi =(F_i^{+}+F_i^{-})+k_\Phi 2c^{-2}\left(F^+_m
\overset\circ V_m^{1\Phi }{\overset{\circ }{V}}_i^{1\Phi }+F_m^{-}{%
\overset{\circ }{V}}_m^{2\Phi }{\overset{\circ}{V}}_i^{2\Phi }\right), \\[5mm] 
\partial _k\tilde{\tau}_{ik}^\Phi =(F_i^{+}+F_i^{-})+\kappa _\Phi 
2c^{-2}\left(F_m^+{\overset{\circ }{V}}_m^{1\Phi }{\overset{\circ }{V}}_i^{1\Phi
}+F_m^{-}{\overset{\circ }{V}}_m^{2\Phi }{\overset{\circ}{V}}_i^{2\Phi }\right).
\end{array}
\label{23}
\end{equation}
Let us compare these two formulas with the equation for the energy-momentum
tensor $T_{ik}$ in electrodynamics: $\partial _kT_{ik}=F_i^{+}+F_i^{-}$.
Since now all tensors ($T_{ik}$, $\tau _{ik}$, $\tilde{\tau}_{ik}$) are
responsible for the distribution and flow of energy, as well as for the momentum
of the same medium $%
\Phi $, one can make 2 different claims:

\subsubsection*{1.}

In an interaction with the mediums $M1$ and $M2$, the vacuum shows its limit
properties by having the characteristics $k_\Phi$ and $\kappa_\Phi$ equal to
zero. An equation for EMT of the vacuum is the well known equation $%
\partial _k T_{ik} =F_i^+ + F_i^-$.

\subsubsection*{2.}

The characteristics $k_\Phi$ and $\kappa_M$ of an interaction of the vacuum
and mediums $M1$ and $M2$ are not zero but sufficiently small, and a right
equation for EMT of the vacuum is equation (\ref{23}) with $k_\Phi\neq0$.

It is clear that, for $k_\Phi \neq 0$, the tensors $\tau _{ik}^\Phi $ and $%
T_{ik}$ will give a different picture of distributions of the flux of energy
and momentum in the vacuum. And if a measurement will have a sufficient
precision to make formula (\ref{23}) more preferable, then there will appear
the quantity $k_\Phi 2c^{-2}(F_m^{+}{\overset{\circ }{V}}_m^{1\Phi }{%
\overset{\circ }{V}}_i^{1\Phi }+F_m^{-}{\overset{\circ }{V}}_m^{2\Phi }{%
\overset{\circ }{V}}_i^{2\Phi })$, and together with this, the mysterious
velocity ${\bf v}^\Phi $ included in the structure of ${\overset{\circ }{V}}%
_i^{1\Phi }$ and ${\overset{\circ }{V}}_i^{2\Phi }$ will also be discovered.

\end{document}